\documentclass[twocolumn,floatfix,prl]{revtex4-2}%
\usepackage{graphicx,color}
\usepackage{amsmath}%
\usepackage{amsfonts}%
\usepackage{amssymb}
\usepackage{bm}

\def\s{{\sigma}}
\def\e{{\epsilon}}
\def\k{{ {\bm k} }}

\def\q{{ {\bm q} }}

\def\0{{ {\bm 0} }}
\def\w{{\omega}}
\def\a{{\alpha}}

\allowdisplaybreaks[4]

\begin{document}
\title{
Robust $T$-Linear Resistivity due to SU(4) Valley and Spin Fluctuation Mechanism in Magic Angle Twisted Bilayer Graphene
}
\author{
Daisuke Inoue, Seiichiro Onari, and Hiroshi Kontani
}
\date{\today }

\begin{abstract}
In the magic angle twisted bilayer graphene (MATBG), 
non-Fermi liquid like transport phenomena are universally observed.
To understand their origin,
we perform the self-consistent analysis of the 
self-energy due to SU(4) valley + spin fluctuations
induced by the electron-electron correlation.
In the SU(4) fluctuation mechanism, the fifteen channels of fluctuations 
contribute additively to the self-energy. 
Therefore, the SU(4) fluctuation mechanism gives much higher
electrical resistance than the spin fluctuation mechanism.
By the same reason,
SU(4) fluctuations of intermediate strength 
provide $T$-linear resistivity down to $\sim1$K.
Interestingly, the $T$-linear resistivity is robustly realized 
for wide range of electron filling, even away from the 
van-Hove filling.
This study provides a strong evidence for the  
importance of electron-electron correlation in MATBG.
\end{abstract}

\address{
Department of Physics, Nagoya University,
Furo-cho, Nagoya 464-8602, Japan. 
}
\sloppy

\maketitle


\section{INTRODUCTION}
Recently, the magic angle twisted bilayer graphene (MATBG) 
has been studied very actively as a platform of 
novel quantum phase transitions 
\cite{Cao1,Cao2,Yankowitz,Lu,Sharpe,Serlin}.
Nearly flatband due to the multi band folding 
with strong electron correlation is formed 
thanks to the honeycomb moir\'{e} superlattice.
The existence of the valley degrees of freedoms and 
the van Hove singularity (vHS) points
leads to exotic strongly correlated electronic states.
The electron filling of the moir\'{e} bands can be
controlled by the gate voltage.
The MATBG is a Dirac semimetal at $n=0$ (charge neutral point),
while Mott insulating state appears at the half filling $|n|=2$.
Various exotic electronic states appear for $|n|\sim2$,
including the unconventional superconducting 
\cite{Cao1,Cao2,Yankowitz,Lu}
and electronic nematic states
\cite{Kerelsky,Choi,Jiang,Cao3}.
Recently, inter-valley coherent order states
with and without time-reversal symmetry attract great attention
\cite{Nuckolls,Kim}


Such exotic multiple phase transitions are believed to be 
caused by strong Coulomb interaction and the
valley+spin degrees of freedoms in the MATBG
\cite{Isobe-super,Chubukov-nematic}.
For example, the nematic bond order is caused by the 
valley+spin fluctuation interference mechanism,
which is described by the Aslamazov-Larkin (AL) vertex correction (VC)
\cite{Onari-TBG,Kontani-rev2,Tazai-LW}.
This mechanism also explains the nematic and smectic states 
in Fe-based superconductors,
\cite{Onari-SCVC,Onari-form,Yamakawa-PRX,Onari-B2g,Chubukov-FeSe,Chubukov-RG,Tsuchiizu-Cu,Onari-Ni}
cuprates, and nickelates
\cite{Tsuchiizu-Cu,Onari-Ni},
and kagome metals
\cite{Tazai-kagome1,Tazai-kagome2,Tazai-kagome3}.
The significance of the AL-VC
has been confirmed by the functional renormalization group (RG) studies 
\cite{Chubukov-FeSe,Chubukov-RG,Tsuchiizu-Cu,Tsuchiizu-Ru1}.
On the other hand, the significance of the electron-phonon interactions 
in the MATBG has been discussed in Refs.
\cite{Sarma1,Sarma2},
and the acoustic phonon can cause the nematic order 
\cite{Fernandes-TBG}.
Thus, the origin and the nature of the strongly correlated electronic states 
in MATBG for $|n|\sim2$ is still uncovered.


To understand the dominant origin of electron correlations,
transport phenomena provides very useful information.
In cuprate and Fe-based superconductors,
non-Fermi-liquid type transport coefficients,
such and the $T$-linear resistivity and 
Curie-Weiss behavior of Hall coefficient ($R_H$),
are naturally explained by the spin fluctuation mechanism
\cite{Kontani-RH,Kontani-MR,Kontani-MR2,Kontani-Nernst,Kontani-rev1}.
The increment of $R_H$ originates from the 
significant memory effect described by the current VC
\cite{Kontani-rev1}.

Interestingly, 
prominent non-Fermi-liquid type transport phenomena
has been universally observed in MATBG.
For example, almost perfect $T$-linear resistivity 
is realized for wide area of $n=\pm (1.0-3.0)$
\cite{Jaoui,Polshyn,Park}.
The Curie-Weiss behavior of $R_H$ is also observed
\cite{Lyu}.
These results are the hallmark of the presence of 
strongly anisotropic quasiparticle scattering.
(In fact, the acoustic phonon scattering mechanism
gives $\rho\propto T^{4}$ at low temperatures
\cite{Sarma1,Sarma2}.)
Thus, non-Fermi-liquid type transport phenomena in MATBG
are significant open problems
to understand the dominant origin and the nature 
of the electron correlation.

In this paper, we study the many-body electronic states in MATBG
in the presence of the SU(4) valley+spin composite fluctuations. 
The self-energy due to the SU(4) fluctuations (${\hat \Sigma}(k)$)
is calculated by employing the fluctuation-exchange (FLEX) approximation.
The obtained resistivity well satisfies the $T$-linear behavior
for $T=1\sim10$K for wide range of $n$.
Large $T$-linear coefficient $a\equiv \rho/T$ is obtained in the present mechanism
due to the contribution of fifteen channel SU(4) fluctuations.
Therefore, the obtained result is quantitatively consistent with experiments.
The present results indicates the development of 
SU(4) valley+spin composite fluctuations in MATBG,
which should be strongly associated with the
exotic multiple phase transitions.

\section{$T$-linear resistivity near the QCP}

In usual Fermi liquids (FLs), the resistivity follows the relations
$\rho=A T^2$ and $A\propto \{N(0)\}^2$ at low temperatures, 
where $N(0)$ is the density-of-states (DOS) at Fermi level 
\cite{Kontani-rev1}.
(Also, the Hall coefficient and the magnetoresistivity in FLs
follow the relations $|R_{\rm H}|\approx 1/en$ and
$\Delta\rho/\rho_0\propto (B_z/\rho)^2$, respectively
\cite{Kontani-rev1}.)
In contrast, $T$-linear resistivity is observed
in two-dimensional (2D) metals near the quantum critical points.
For example, Ce$M$In$_5$ ($M$=Co, Rh)
exhibits non-FL like relationships such as
$\rho\sim T$ and $R_{\rm H}\sim T^{-1}$,
in addition to the modified Kohler's rule 
$(\Delta\rho/\rho_0)\propto (R_{\rm H}/\rho)^2$
\cite{Nakajima1,Nakajima2}.
Similar non-FL transport phenomena are observed
near the nematic quantum critical point (QCP) in $\mathrm{Fe(Se,S)}$
\cite{FeSe1,FeSe2}. 
Furthermore, $T$-linear resistivity appears
in nickerates \cite{Ni1,Ni2}
and cuprates \cite{YBCO,Taillefer}
near the charge-density-wave (CDW) QCPs.

To understand the critical transport phenomena,
the self-consistent renormalization (SCR) theory
\cite{Moriya-rev1},
the renormalization group theory
\cite{Hertz,Millis},
spin-fermion model analysis
\cite{Rice,Pines,Millis,Abanov}
have been performed.
In these theories, strong quasiparticle scattering rate
$\gamma_\k={\rm Im}\Sigma_\k^A(0)$ due to quantum fluctuations gives rise to the 
non-FL resistivity $\rho\propto T^n$ with $n<2$ near the QCP.
($n=1 \ [4/3]$ in 2D metals with the antiferro (AF) [ferro] fluctuations
according to Ref. \cite{Moriya-rev1}.)
More detailed analyses are explained in Ref. 
\cite{Abanov}.

It is noteworthy that the current VC 
plays significant roles in both $R_{\rm H} \ (\propto T^{-1})$ 
and $\Delta\rho/\rho_0 \ (\propto T^{-2}\rho^{-2})$,
in addition to the self-energy 
\cite{Kontani-rev1}.
The modified Kohler's rule 
$(\Delta\rho/\rho_0)\propto (R_{\rm H}/\rho)^2$
observed in CeMIn$_5$ and the Fe(Se,S)
is naturally explained by considering the current VC 
\cite{Kontani-rev1}.

Here, we concentrate on the $T$-dependence of the resistivity,
where the current VC is not essential.
In the SCR theory and the spin-fermion model,
the dynamical AF susceptibility is assumed as
\begin{eqnarray}
\chi^{\rm AF}(\q,\w)=\frac{\chi^{\rm AF}_0}{1+\xi^2(\q-{\bm Q})^2-i\w/\w_{\rm AF}}
\end{eqnarray}
where $\xi$ is the AF correlation length and 
${\bm Q}$ is the AF wavevector.
$\w_{\rm AF}$ is the energy scale of the AF fluctuations
and $\chi^{\rm AF}_0=\chi^{\rm AF}({\bm Q},0)$:
They are scaled as $\w_{\rm AF}\propto \xi^{-2}$
and $\chi^{\rm AF}\propto\xi^2$
\cite{Moriya-rev1,Kontani-rev1,Rice,Pines}.
The relation $\xi^2\propto (T-T_0)^{-1}$ is satisfied for 
wide parameter range, and $T_0=0$ at the QCP.
In the SCR theory, when $\w_{\rm AF}\lesssim T$,
the resistivity is approximately given as
$\rho\sim \sum_\k\gamma_\k \sim T^2 \sum_{\k\k'} \rho_{\k'}(0){\rm Im}\chi^{\rm AF}(\k-\k',\w)/\w|_{\w=0}\sim T^2\xi^{4-d}$,
where $\rho_\k(\w)={\rm Im}G^A_\k(\w)/\pi$
\cite{Moriya-rev1,Kontani-rev1}.
Thus, the $T$-linear resistivity appears when $T_0\sim0$.

In various two-dimensional Hubbard models,
the relation $\rho\propto T$ is reproduced 
based on the FLEX approximation \cite{FLEX1,FLEX2,FLEX3},
because the relation $\xi^2\propto T^{-1}$ is well satisfied 
for $U\sim W_{\rm band}$.
(Note that the relation $\xi^2 \propto (1-\a)^{-1}$ holds,
where $\a$ is the Stoner factor given by the FLEX approximation.)
Importantly, the relation $\xi^2\ll\infty$ for $T>0$
 is always satisfied 
by the FLEX approximation for two-dimensional systems
because the FLEX approximation satisfies the Mermin-Wagner theorem,
as analytically proved in the Appendix of Ref. 
\cite{MW}.

In Ref. \cite{Onari-TBG},
the present authors studied realistic Hubbard model for MATBG 
\cite{Koshino}
based on the RPA, and derived the development of the 
SU(4) valley+spin composite fluctuations.
The nematic bond-order is caused by the interference between
SU(4) fluctuations
\cite{Onari-TBG}.
In this paper, we study the same MATBG model
based on the FLEX approximation, 
where the self-energy is calculated self-consistently.

Thanks to the self-energy, the $T$-linear resistivity is realized 
for wide parameter range.
Interestingly, 
the $T$-linear resistivity is realized even when the 
system is far from the SU(4) QCP 
so that $\xi^2T^2$ decreases at low temperatures.
The present result indicates that 
the $T$-linear resistivity in MATBG
originates from the combination between the 
moderate SU(4) fluctuations and the characteristic 
band structure with the vHS points.
Importantly,
the $T$-linear coefficient $a=\rho/T$ is large in the present
fifteen-channel SU(4) fluctuation mechanism, 
compared with the conventional three-channel 
SU(2) spin fluctuation mechanism.
Consistently, the observed $a$ is rather large in MATBG
\cite{Jaoui,Polshyn}.






 



\section{formulation}
\begin{figure}[!htb]
\includegraphics[width=.99\linewidth]{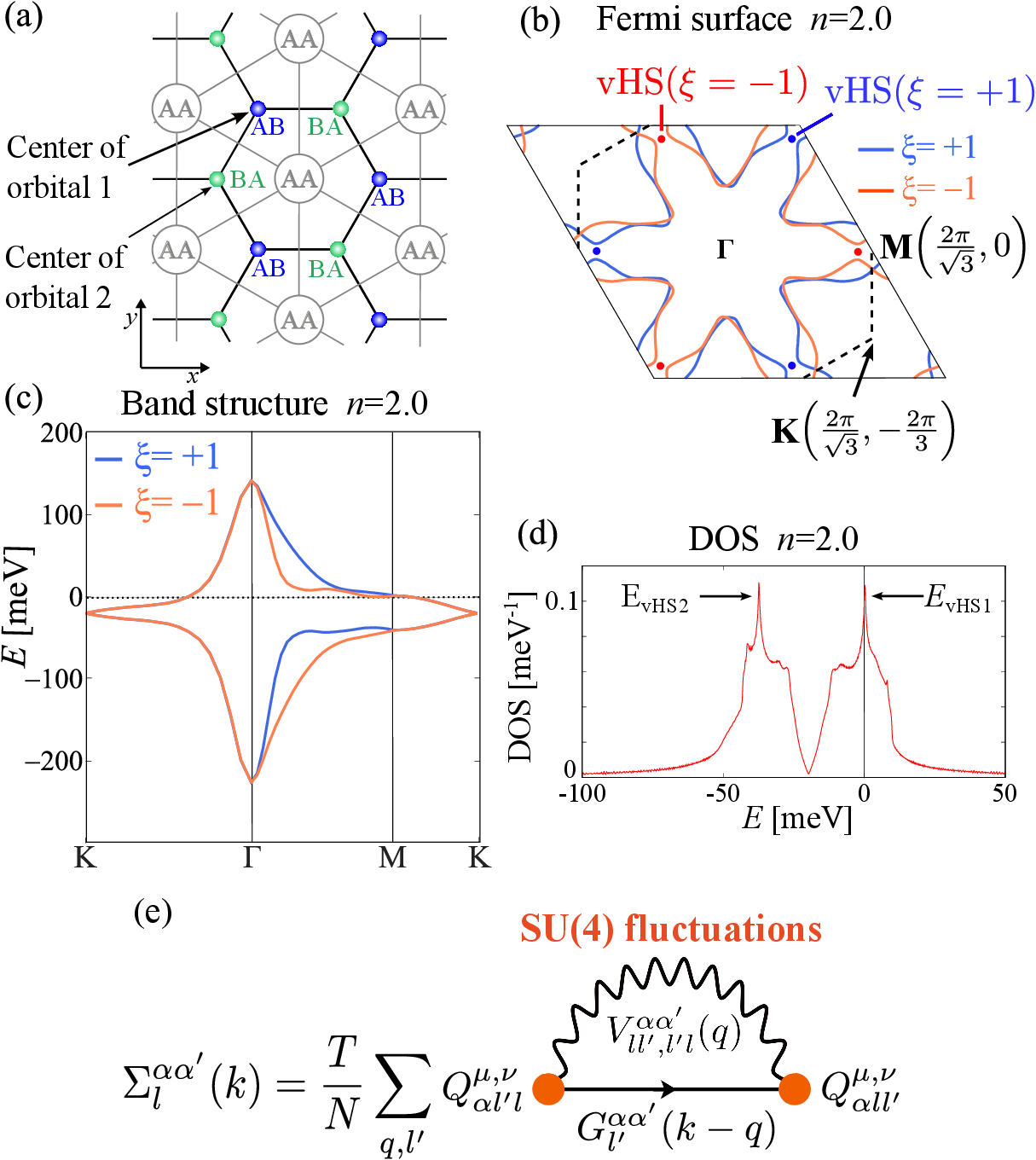}
\caption{
(a) Lattice structure of the MATBG model. Wannier orbitals 1 and 2 
are centerd at AB (blue) and BA (green) sublattices, respectively.
(b) FSs for $n=2.0$ and the vHS points, 
where orange (blue) lines and dots correspond to the valley for $\xi = +1\ (-1)$
, respectively.
(c) Band structure of the MATBG model.
(d) DOS for $n=2.0$, which has vHS points at $E_{\mathrm{vHS}1}$
and $E_{\mathrm{vHS}2}$.
(e) Feynman diagram of the self-energy in the FLEX approximation.
}
\label{fig:fig1}
\end{figure}

Here, we analyze the following multiorbital model for MATBG
studied in Ref. \cite{Koshino,Onari-TBG}:
\begin{eqnarray}
H^0= \sum_{\k,\a\a'l}c_{\k,\a l}^\dagger h^0_{\a\a'l}(\k) c_{\k,\a' l},
\label{eqn:H0}
\end{eqnarray}
where $\k=(k_x,k_y)$, $l=(\rho,\xi)$, $\rho$ and $\xi$ represent spin
and valley indices, respectively.
%
Here, $\a=A \ (B)$ which represents a sublattice AB (BA) 
is the center of Wannier orbital 1 (2)
in Fig. \ref{fig:fig1} (a).
Also, the valley index $\xi=\pm1$ correspond to the angular momentum.
This model Hamiltonian is based on the
first-principles tight-binding model in Ref. \cite{Koshino},
and we modified the hopping integrals according to Ref. \cite{Onari-TBG}.

The Fermi surface (FS) of this model at $n=2.0$ 
is shown in Fig. \ref{fig:fig1} (b).
Here, two FSs are labeled as $\xi=+1$ and $\xi=-1$
because $H^0$ is diagonal with respect to the valley.
Six vHS points are shown in Fig. \ref{fig:fig1} (b).
The band structure and total DOS are given
 in Fig. \ref{fig:fig1} (c) and Fig. \ref{fig:fig1} (d),
 respectively.
 Energy gap between the two vHS energies $E_{\mathrm{vHS1}}-E_{\mathrm{vHS2}}\sim 50$ meV
 corresponds to the effective bandwidth, which is consistent with
 the STM measurement \cite{Kerelsky}.


\begin{figure}[htb]
	\includegraphics[width=.99\linewidth]{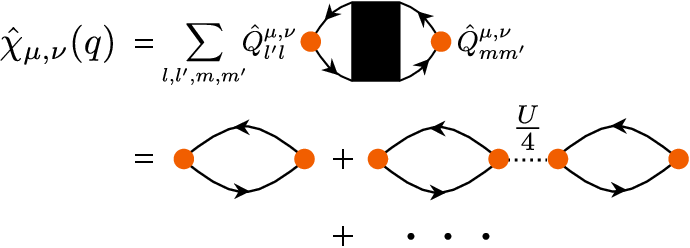}
	\caption{Diagram of the SU(4) susceptibility
	$\hat{\chi}_{\mu,\nu}(q) \ [(\mu,\nu)\neq(0,0)]$.
	}
	\label{fig:fig_diagram}
	\end{figure}

The $2\times 2$ matrix Green function with respect to
the sublattices (A,B) is given as
\begin{eqnarray}
\hat{G}_{l}(k)=\left[(i\e_n-\mu){\hat 1}-{\hat h_l}^0(\k)-{\hat{\Sigma}}_l(k)\right]^{-1},
\label{eqn:G}
\end{eqnarray}
where $k\equiv(\k,i\e_n)$, $\e_n=(2n+1)\pi T$ and
$\mu$ is the chemical potential, and ${\hat \Sigma}_l(k)$
is the self-energy.

In MATBG, the intra- and inter-valley on-site Coulomb interactions
are exactly the same ($U=U'$) \cite{Koshino}.
Also, the inter-valley exchange interaction
$J$ is very small ($J/U\ll 1$) \cite{Koshino,Klug}, 
therefore we set $J=0$.
Then, the Coulomb interaction term is given as
\begin{align}
H_U&=& \frac{U}{2}\sum_{i,\a\xi}
\left( \sum_{\rho\rho'}n_{i,\a\rho\xi}n_{i,\a\rho'\bar{\xi}}
+\sum_{\rho}n_{i,\a\rho\xi}n_{i,\a\bar{\rho}\xi} \right),
\label{eqn:HU1}
\end{align}
where $i$ is the unit cell index.
$n_{i,\a\rho\xi}$ is the electron number operator with spin $\rho$
and valley $\xi$ at $\a$ spot.
Using SU(4) operators in Eq. \ref{eqn:O-def},
$H_U$ is expressed as \cite{Onari-TBG}
\begin{eqnarray}
H_U&=&\frac{U}{16} \sum_{i,\a}\left[ -\sum_{\mu,\nu}({O}^{i,\a}_{\mu,\nu})^2 +4({O}^{i,\a}_{0,0})^2\right],
\label{eqn:HU2}
\\
{O}^{i,\a}_{\mu,\nu}&=& \sum_{ll'} Q^{\mu,\nu}_{\alpha ll'}c_{i,\a l}^\dagger c_{i,\a l'},
\label{eqn:O-def}
\end{eqnarray}
where $\mu,\nu=0\sim3$ and 
${Q}^{\mu,\nu}_{\a ll'}=({\hat \s}_\mu\otimes {\hat \tau}_\nu)_{ll'}$.
Here, ${\hat \s}_m$ (${\hat \tau}_m$) for $m=1,2,3$ 
is Pauli matrix for the spin-channel with $\rho=\pm1$
(valley-channel with $\xi=\pm1$).
${\hat \sigma}_0$ and ${\hat \tau}_0$ are the identity matrices.
The Coulomb interaction $H_U$ in Eq. (\ref{eqn:HU2}) apparently
possesses the SU(4) symmetry.
Note that similar multipolar decomposion of 
the Coulomb interaction has been used
in the strong heavy fermion systems in Refs. \cite{Tazai-multipole1,Tazai-multipole2,Tazai-multipole3}.

Here, we examine the 
SU(4) susceptibility given as 
\begin{align}
	\chi_{\mu, \nu;\mu',\nu'}^{\a\a'}(\q,i\w_l)=
\int_0^\beta d\tau \langle {O}^{\a}_{\mu,\nu}(\q,\tau){O}^{\a'}_{\mu', \nu'}(-\q,0)\rangle
e^{i\w_l\tau},
\label{eqn:chi1}
\end{align}
where $q\equiv(\q,\w_l)$ and $\w_l=2l \pi T$.
In the present calculations, we consider only
diagonal channels with respect to $(\mu,\nu)$, $\chi_{\mu, \nu;\mu, \nu}^{\a \a'}$, because
off-diagonal channels $\chi_{\mu,\nu;\mu'\nu'}^{\a \a'}\  [(\mu',\nu') \neq (\mu,\nu)]$
are exactly zero or very small.
Then, diagonal channel $\chi_{\mu,\nu}^{\a \a'}(q)$ except for $(\mu,\nu)=(0,0)$
is expressed as

\begin{equation}
	\begin{split}
		\hat{\chi}_{\mu,\nu}(q) &= \hat{\chi}^0_{\mu,\nu}(q) +\frac{U}{4}\hat{\chi}^0_{\mu,\nu}(q)\hat{\chi}^0_{\mu,\nu}(q) + \cdots \\
    &=\hat{\chi}^0_{\mu,\nu}(q)\left(\hat{1}-\frac{U}{4}\hat{\chi}^0_{\mu,\nu}(q)\right)^{-1}, \\
\label{eqn:chi2}
	\end{split}
\end{equation}
\begin{align}
	\chi^{0;\a \a'}_{\mu,\nu}(q) &=-\frac{T}{N}\sum_{k,ll'}{Q}^{\mu,\nu}_{\a l'l}{Q}^{\mu,\nu}_{\a' l l'}
	G_{l}^{\a \a'}(k+q)G_{l'}^{\a' \a}(k).
	\label{eqn:chi0}
\end{align}
Figure \ref{fig:fig_diagram} shows the diagrammatic expression in Eq. (\ref{eqn:chi2}).
Here, $\hat{\chi}_{m, 0}(q)$
represents the spin susceptibility, 
$\hat{\chi}_{0, m}(q)$ represents the valley susceptibility,
and $\hat{\chi}_{m, n}(q)$ represents the
susceptibility of the "spin-valley quadrupole order".
Also, the local charge susceptibility $\hat{\chi}_{0, 0}(q)$ is expressed
as
\begin{align}
	\hat{\chi}_{0, 0}(q) =
    \hat{\chi}^0_{0, 0}(q)\left(\hat{1}+\frac{3U}{4}\hat{\chi}^0_{0, 0}(q)\right)^{-1},
	\label{eqn:chi-00}
\end{align}
which is suppressed by $U$.

In the FLEX approximation, the self-energy and the effective interaction
are given as
\begin{align}
		\Sigma_{l}^{\a \a'}(k) &= \frac{T}{N}\sum_{q,l'}
		G_{l'}^{\a \a'}(k-q)V^{\a \a'}_{ll',l'l}(q), 
	\label{eqn:self}\\
	V^{\a \a'}_{l l', l'l}(q)&=\left(\frac{U}{4}\right)^2 \sum_{\substack{\mu,\nu \\\neq (0,0)}}
	Q^{\mu,\nu}_{\alpha ll'}\chi_{\mu,\nu}^{\a \a'}(q)Q^{\mu,\nu}_{\a' l' l} \nonumber \\
	&\ \ \ \ \ + \left( \frac{3U}{4} \right)^2 Q^{0,0}_{\alpha ll'}\chi_{0,0}^{\a \a'}(q)Q^{0,0}_{\a' l' l}.
	\label{eqn:V}
\end{align}
Here, we solve Eqs. (\ref{eqn:chi2})-(\ref{eqn:V}), 
self-consistently.
Note that the double-counting
$U^2$ terms in Eqs. (\ref{eqn:A-self}) are subtracted properly.
In the present numerical study, we
use $108\times108\  \bm{k}$ meshes and 2048
Matsubara frequencies.

In the case of SU(4) symmetry limit, 
the Green function $\hat{G}_l(k)$ is 
independent of the spin and valley.
Then, it is allowed to replace $\hat{G}_l(k)$
in Eq. (\ref{eqn:chi0}) with
$\hat{G}_{\mathrm{av}}(k)\equiv 1/4\sum_l \hat{G}_{l}(k)$.
Therefore,
the irreducible susceptibility in the SU(4) symmetry limit
is approximately simplified as
\begin{align}
	\hat{\chi}_{\mu,\nu}^{0}(q) \approx 4\hat{\chi}_{\mathrm{av}}^{0}(q),
\end{align}
where $\chi^{0;\a \a'}_{\mathrm{av}}(q)\equiv -\frac{T}{N}\sum_{k}G_{\mathrm{av}}^{\a \a'}(k+q)G_{\mathrm{av}}^{\a' \a}(k)$.
Here, we used the relation $\sum_{ll'}{Q}^{\mu,\nu}_{\a l' l}{Q}^{\mu,\nu}_{\a l l'} = 4$
for all $\mu,\nu$.
Also, the SU(4) susceptibility except for $(\mu,\nu) = (0,0)$ 
in Eq. (\ref{eqn:chi2})
and the self-energy in Eq. (\ref{eqn:self})
in the SU(4) symmetry limit is given as
\begin{align}
	\hat{\chi}_{\mu,\nu}(q) &\approx 4\hat{\chi}_{\mathrm{av}}(q) \\ \nonumber
	&\equiv 4\hat{\chi}_{\mathrm{av}}^0(q)\left(\hat{1}-U\hat{\chi}^0_{\mathrm{av}}(q)\right)^{-1}, \\
	\Sigma^{\a \a'}(k) &\approx \frac{T}{N}\sum_{q}\frac{15}{4}U^2 G^{\a \a'}_{\mathrm{av}}(k-q)\chi_{\mathrm{av}}^{\a \a'}(q).
	\label{eqn:self_SU(4)}
\end{align}
Eq. (\ref{eqn:self_SU(4)}) indicates that 
the self-energy per orbital in this system
develops easier than the systems
which are considerd spin or charge fluctuations,
due to the multi-channel SU(4) fluctuations.

In the presence of the off-site Coulomb interaction
between $(i,\a)$ and $(j,\a')$, $v_{i\a,j\a'}$,
the interaction Hamiltonian is given as
\begin{equation}
	\begin{split}
		H_v&= \sum_{ij,\a \a' l l'}
v_{i\a,j\a'}c_{i,\a l}^\dagger c_{i,\a l}c_{j,\a' l'}^\dagger c_{j,\a' l'}\\
&= \sum_{ij,\a\a'}v_{i\a,j\a'}O_{0,0}^{i,\alpha}O_{0,0}^{j,\alpha'}
\label{eqn:off-site}
	\end{split}
\end{equation}
Then, the effect of off-site Coulomb interaction in the FLEX approximation 
is simply given by replacing $(3U/4)^2$ in Eq. (\ref{eqn:V})
with $(3U/4 + 2v_{\a \a'}(\q))^2$.
Here, $v_{\a \a'}$ is the Fourier transform of
$v_{i \a,j\a'}$.

Present formulation using the Coulomb interaction expressd by the SU(4) operator
is equivalent to the conventional formulation using the
Coulomb interaction expressed by the spin and charge channels.
We explain the correspondence with the previous multiorbital
FLEX approximation formalism in Appendix A.

We obtain the resistivity $\rho=1/\sigma_{xx}$
based on the Kubo formula.
$\sigma_{xx}$ is given by
\begin{align}
	\sigma_{xx} = e^2\sum_{\bm{k},\a \xi} \int \frac{d\omega}{\pi}\left( -\frac{\partial f}{\partial \omega}\right)
	|G^{\a}_{\xi}(\bm{k},\omega)|^2 (v^{\a}_{\xi;x}(\bm{k},\omega))^2,
	\label{eqn:conductivity}
\end{align}
where $v^{\a}_{\xi;x}(\bm{k},\omega)=\partial(\epsilon^{\a}_{\xi}
+ \mathrm{Re}\Sigma^{\a}_{\xi}(\bm{k},\omega))/\partial k_x$
is the quasiparticle velocity, and
$f=1/(1+e^{(\omega-\mu)/T})$.
Here, $\a$ and $\xi$ denote the sublattice and valley, 
respectively.
The self-energy $\Sigma^{\a}_{\xi}(\bm{k},\omega)$
is obtained by the analytic continuation of Eq. (\ref{eqn:self})
using Pade approximation.
Eq. (\ref{eqn:conductivity}) is transformed by using the relation
$|G(\bm{k},\omega)|^2=\pi \rho_{\bm{k}}(\omega)/\gamma_{\bm{k}}(\omega)$, 
where $\rho_{\bm{k}}(\omega)$ is the quasiparticle weight.
In the present study, we drop the current vertex 
corrections (CVC), which are necessary to describe
 the Umklapp scatterings. As explained in Ref. \cite{Kontani-rev1},
 the T-liner resistivity near the QCP is altered by the CVC only
 quantitatively, although the CVC is essential for the quantum
 critical behavior of the Hall coefficient. For this reason,
 the CVC is ignored for simplicity in the present study.

Finally, we comment on the 
"topological obstruction” of the present tight-binding model.
It is known that the 
effective tight-binding model with well-localized Wannier orbitals 
has a lack of symmetry captured within the continuum theory 
in MATBG, which is so-called “topological obstruction”
 \cite{Senthil1,Senthil2,Meng}.
 To avoid the obstructions,
  some ways such as introduction of
  an assisted-hopping interaction are proposed in the previous studies
  \cite{Vafek,Meng, Meng2}.
 Indeed, our model has no $C_2T$ symmetry,
 where $C_2$ and $T$ represent twofold symmetry with respect
 to the z axis and time reversal symmetry, respectively.
 However, existence of the $C_2T$ symmetry becomes important
 for the electronic states at charge neutral point ($n=0$)
 \cite{Meng,Meng2,Lee}. On the other hand, we study the 
 transport phenomena for $n=1-3$,
 which is a good metal with large Fermi surfaces
 and the Dirac points are far away from the Fermi
 level. Therefore, the present tight-binding model is
 suitable for analyzing the non-Fermi liquid behavior in MATBG.

\begin{figure}[htb]
	\includegraphics[width=.99\linewidth]{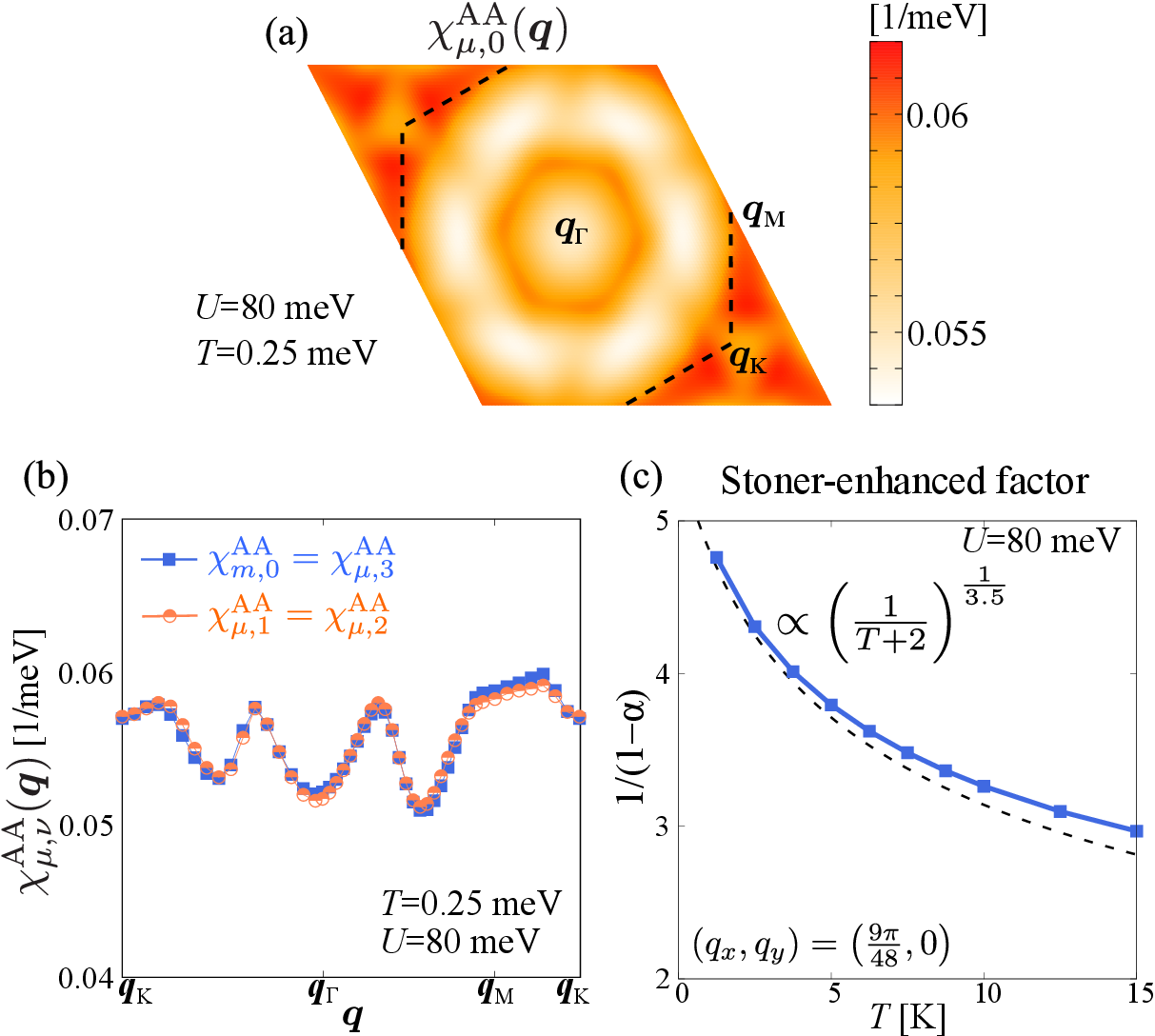}
	\caption{
	(a) $\bm{q}$-dependences of the spin susceptibility 
	$\chi_{\mu,0}^{\mathrm{AA}}(\bm{q},\omega=0)$. 
	(b) $\chi_{\mu, \nu}^{\mathrm{AA}}(\bm{q})$ obtained by the FLEX approximation.
	(c) $T$-dependence of the Stoner-enhanced factor $\a$.
	}
	\label{fig:fig2}
	\end{figure}
\section{numerical result}

Hereafter, we mainly study the case of
$n=2.0$, where the Fermi level is close to vHS energy.
We consider only the on-site Coulomb interaction
unless otherwise noted. 
 Figure \ref{fig:fig2} (a) and \ref{fig:fig2} (b) show 
the SU(4) susceptibility $\chi_{\mu,\nu}^{\mathrm{AA}}(\bm{q})$, $\mu,\nu=0\sim3$
 [$(\mu,\nu) \neq (0,0)$].
 In the present calculation,
 $\chi_{\mu,\nu}^{\mathrm{AA}} (\bm{q})\simeq \chi_{\mu,\nu}^{\mathrm{BB}}(\bm{q})
 > \chi_{\mu,\nu}^{\mathrm{AB}}(\bm{q}) \simeq \chi_{\mu,\nu}^{\mathrm{BA}}(\bm{q})$
 is satisfied.
 $\hat{\chi}_{\mu,\nu}(\bm{q})$ include not only the spin fluctuations
 but also, valley and valley$+$spin composite fluctuations.
 The fifteen components of
 $\hat{\chi}_{\mu,\nu}(\bm{q})$ take very similar values 
 by reflecting the SU(4) symmetry
 Coulomb interaction in Eq. (\ref{eqn:HU2}).
 As shown in Fig.\ref{fig:fig2} (b), 
 seven components with $(\mu,\nu) = (m,0),\ (\mu,3)$ 
 are exactly equivalent, 
 and eight components with $(\mu,\nu) = (\mu,1),\ (\mu,2)$ are also equivalent 
 , where $m=1\sim3$.
 In the present MATBG model given in Eq. (\ref{eqn:H0}),
 FSs are different with respect to the
 valley index as shown in Fig. \ref{fig:fig1}(b), but 
 the difference is very small.
 Therefore, the system possesses approximate
 SU(4) symmetry and the fifteen channels
 of $\hat{\chi}_{\mu,\nu}$ equally develop.
 Note that $\hat{\chi}_{0,0}$ is much smaller value than
 that in other channels
 ($\hat{\chi}_{0,0} \thicksim 1/10\hat{\chi}_{\mu,\nu}$).
 $\chi_{\mu,\nu}^{\mathrm{AA}}(\bm{q})$ develops around
 the nesting vector that connects the two vHS points.
 The Stoner factor $\alpha$ is defined as the largest eigenvalue
 of $U \hat{\chi}^0_{\mu,\nu}(\bm{q},0)/4\approx U \hat{\chi}^0_{\mathrm{av}}(\bm{q},0)$.
 It represents the SU(4) fluctuation strength.
 Figure \ref{fig:fig2}(c) shows the $T$-dependence of 
 the Stoner-enhanced factor.
 According to the spin fluctuation theory \cite{Moriya-rev1},
 the relation $1/(1-\alpha) \propto 1/T$ is satisfied due to
 the development of $\a$ at low temperatures,
 and this relation gives rise to
 the $T$-linear resistivity.
 On the other hand, in the present calculations, 
 $\alpha \lesssim  0.8$ and $1/(1-\alpha) \propto (1/T+2)^{1/3.5}$
 indicate an interesting deviation from the
  conventional spin fluctuation theory in MATBG.

\begin{figure}[htb]
\includegraphics[width=.99\linewidth]{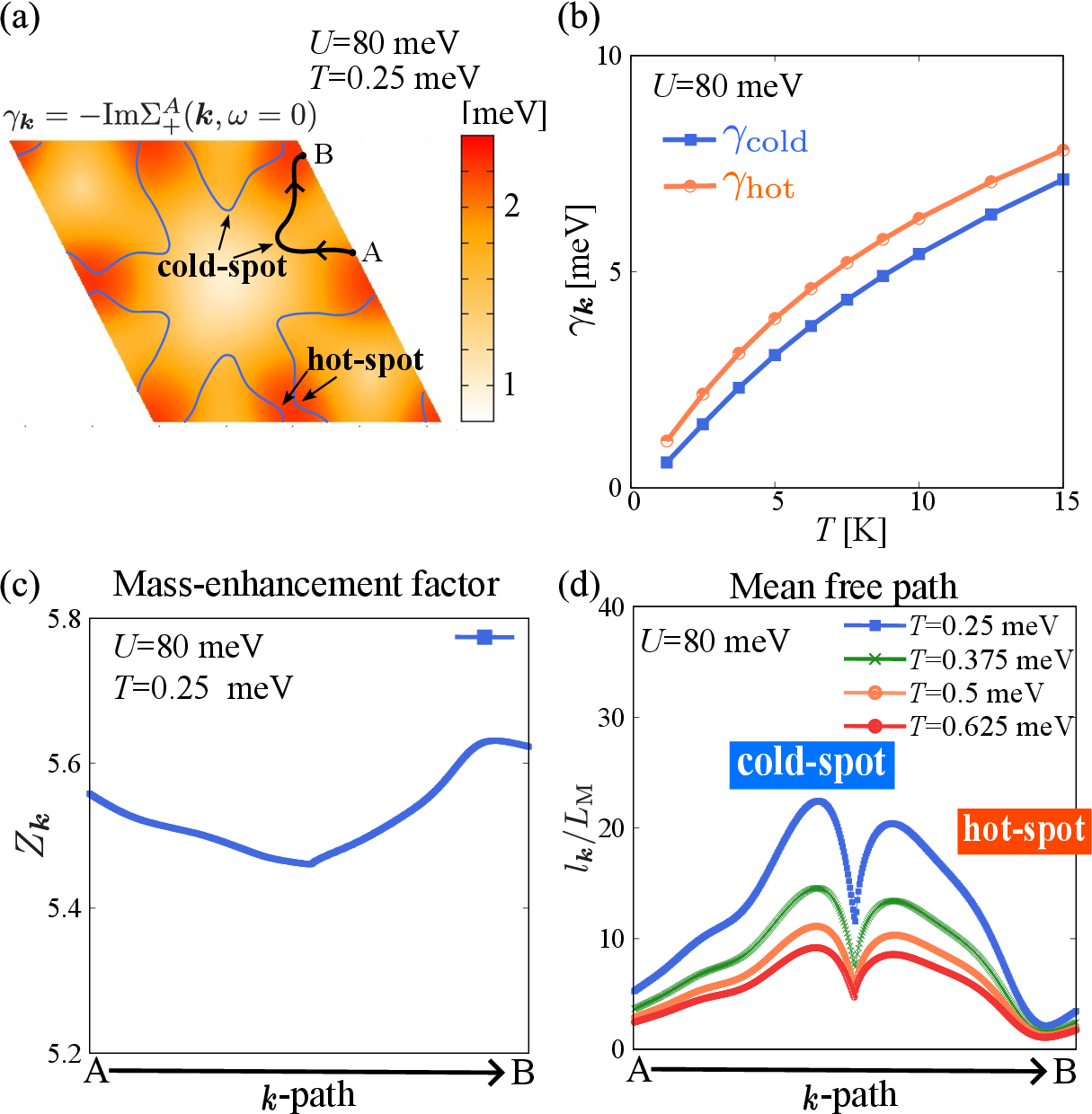}
\caption{
(a) $\bm{k}$-dependences of $\gamma_{\bm{k}}$ for $\a=A,\xi = +1$
, where blue line represents the FS.
$k$-path is defined by the path from A to B on FS.
(b) $T$-dependence of $\gamma_{\bm{k}}$ at cold ($\gamma_{\mathrm{cold}}$)
and hot ($\gamma_{\mathrm{cold}}$) spot.
$\bm{k}$-dependences of the (c) mass-enhancement factor 
and (d) mean free path along the $k$-path shown in (a).
}
\label{fig:fig3}
\end{figure}
Here, we show the self-energy $\Sigma^{\a}_{\xi}(\bm{k},\omega)$ 
obtained by the FLEX approximation.
The self-energy gives the quasiparticle damping rate
and the mass-enhancement factor.
The quasiparticle damping rate $\gamma_{\bm{k}}$ is defined as
$\gamma_{\bm{k}}=-\mathrm{Im}\Sigma^{\mathrm{A}}_+(\bm{k},0)\simeq-\mathrm{Im}\Sigma^{\mathrm{B}}_+(\bm{k},0)$.
Figure \ref{fig:fig3}(a) shows the
$\bm{q}$-dependences of the $\gamma_{\bm{k}}$
 due to the SU(4) fluctuations.
 There are hot (cold) spots, where $\gamma_{\bm{k}}$
 takes maximum (minimum) value.
 The hot spots exist near the vHS points.
 Fig. \ref{fig:fig3}(b) shows the $T$-dependence
 of $\gamma_{\bm{k}}$ at hot and cold spots
 ($\gamma_{\mathrm{hot}}, \gamma_{\mathrm{cold}}$).
 The $T$-dependence of $\rho$ follows roughly
 that of $\gamma_{\mathrm{cold}}$.
 In our calculations, although the fluctuation per one channel is
 weak ($\a \lesssim 0.8$) away from the SU(4) QCP, $\gamma_{\mathrm{cold}} \propto T$ 
 is realized at low temperatures
 owing to the fifteen-channel SU(4) fluctuations. 
 In the present study, we discuss the resistivity for
  $T > 1$K because the present numerical results using
   108 $\bm{k}$-meshes and 2048 Matsubara numbers 
   become less accurate for $T\lesssim 1$K.

 The mass-enhancement factor $Z_{\bm{k}}$ and
 the mean free path $l_{\bm{k}}$ are given as
\begin{eqnarray}
	Z_{\bm{k}} &=& 1-\left. \frac{\partial \mathrm{Re}\Sigma_{+}^{\mathrm{A}}(\bm{k},\omega)}{\partial \omega}\right|_{\omega=0}, \\
	l_{\bm{k}} &=& \left\lvert \frac{\bm{v}^{\a}_{\xi}(\bm{k},0)}{\gamma_{\bm{k}}}\right\rvert, 
\end{eqnarray}
where $\bm{v}^{\a}_{\xi}(\bm{k},0)$ is the quasiparticle velocity.
Fig. \ref{fig:fig3}(c) shows the mass-enhancement factor
$Z_{\bm{k}} = m^*/m$
along the $\bm{k}$-path on the FS shown 
in Fig. \ref{fig:fig3}(a),
where $m$ and $m^{*}$ are the bare electron mass
and the effective mass, respectively.
The obtained $Z_{\bm{k}}>5$ indicates that this system is
in the strongly correlated region.
Fig. \ref{fig:fig3}(d) shows the obtained $l_{\bm{k}}$ 
devided by the moir\'{e} superlattice constant $L_{\mathrm{M}}$.
$l_{\bm{k}}$ on the FS is longer than $L_{\mathrm{M}}$, particularly
$l_{\mathrm{cold}} \sim 20 L_{\mathrm{M}}$ at $T\approx 3K$,
 where $l_{\mathrm{cold}}$ is $l_{\bm{k}}$ at cold spots.
Such long $l_{\bm{k}}$ and large $Z_{\bm{k}}$
guarantee that the strongly correlated Fermi liquid state
is realized in this system.
Also, long $l_{\bm{k}}$ is observed experimentally 
at low temperatures ($T\lesssim 10$K)
\cite{Jaoui}. This suggests that
 the Fermi liquid picture holds well, and
 the FLEX approximation is appropriate
 for the analysis of the transport phenomena in MATBG.
 By the FLEX approximation, the quantum and 
 thermal fluctuations are properly considered.
 Therefore, the FLEX method  has great advantages for studying the
 critical phenomena due to the SU(4) fluctuations
 in comparison with several strong-coupling theories
 such as DQMC and DMFT 
 \cite{Isobe-super,Chubukov-nematic,Meng,Meng2,Laissa,Lee}.

\begin{figure}[!htb]
\includegraphics[width=.99\linewidth]{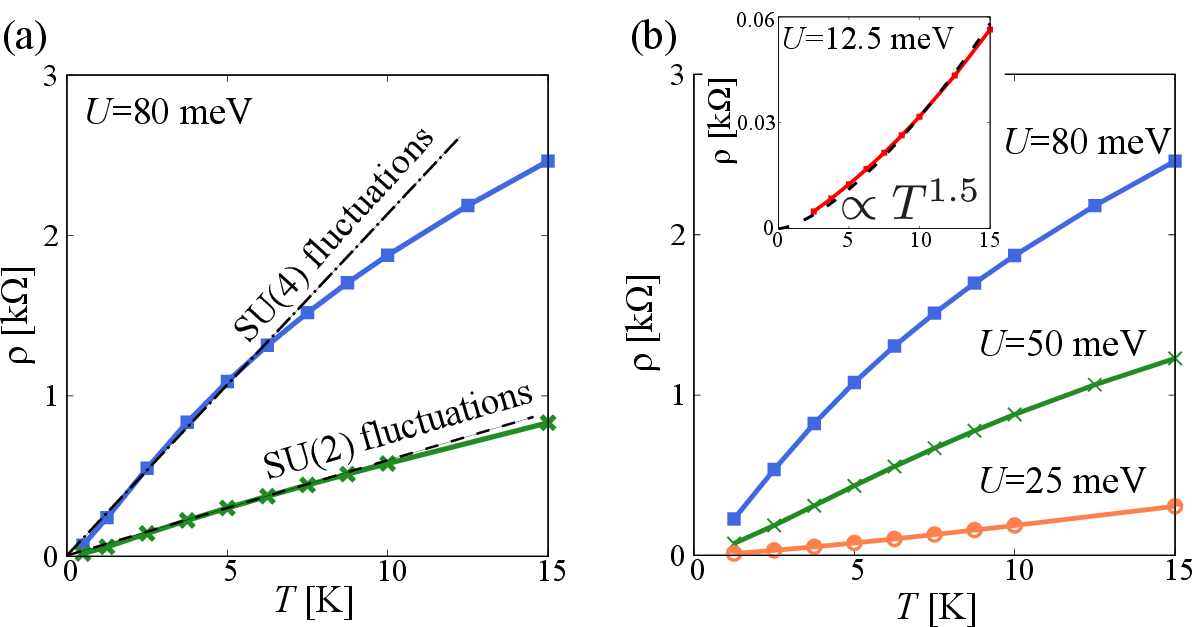}
\caption{
(a) $T$-dependence of $\rho$ obtained by FLEX approximation with 
SU(4) fluctuations and, that with SU(2) fluctuations 
for $U=80$ mev.
(b) $T$-dependence of $\rho$ 
for $U=12.5, 25, 50, 80$ mev. 
}
\label{fig:fig4}
\end{figure}
Figure \ref{fig:fig4}(a) shows the resistivity $\rho$
 obtained by the FLEX approximation
 (blue line) due to the SU(4) fluctuations .
 $\rho \propto T$ is satisfied at low temperatures,
 which is quantitatively consistent with experimental
 results in Refs. \cite{Jaoui,Polshyn,Park}.
 The green line in Fig. \ref{fig:fig4}(a) shows
 $\rho$ given by the FLEX approximation with including
 only spin fluctuations (SU(2) fluctuations).
 The $T$-linear coefficient $a = \rho/T$
 due to the SU(4) fluctuations and that due to the only SU(2) fluctuations
  are $a \sim 0.2$ and $a \sim 0.06$, respectively.
In experimental results \cite{Jaoui,Polshyn}, the observed $T$-linear coefficient
 is larger than 0.1, thus our result considering the SU(4) fluctuations
 is consistent with the observations.
 On the other hand, the $T$-linear coefficient $a$
 due to only the SU(2) fluctuations is very small.
 Therefore, the fifteen-channel SU(4) fluctuations
 are significant for the large $a$.
 We stress that the power $m$ in $\rho = aT^m$ decreases less than 1
 at high temperature.
 This behavior is consistent with some experimental results 
 \cite{Jaoui,Polshyn,Park},
 and realized in previous theoretical study based on
 the FLEX approximation
 \cite{Kontani-RH,Kontani-rev1}.
 Fig. \ref{fig:fig4}(b) shows the obtained $U$-dependence of $\rho$.
 The power $m$ increases as the Coulomb interaction
 becomes weak.
 This behavior indicates that the system approaches
 the Fermi liquid state ($\rho \propto T^2$)
 as $U\rightarrow0$.
 Thus, the  $T$-linear resistivity
 originates from the strong electron-electron correlation effect.
 Here, the power $m$ is smaller than 1.5 even
 when $U=12.5$ meV.
 As we discuss in the Appendix B,
 the power $m$ is smaller than 2 when
 the vHS points near the FS even
 when $U \ll W_{\mathrm{band}}$.

\begin{figure}[!htb]
	\includegraphics[width=.99\linewidth]{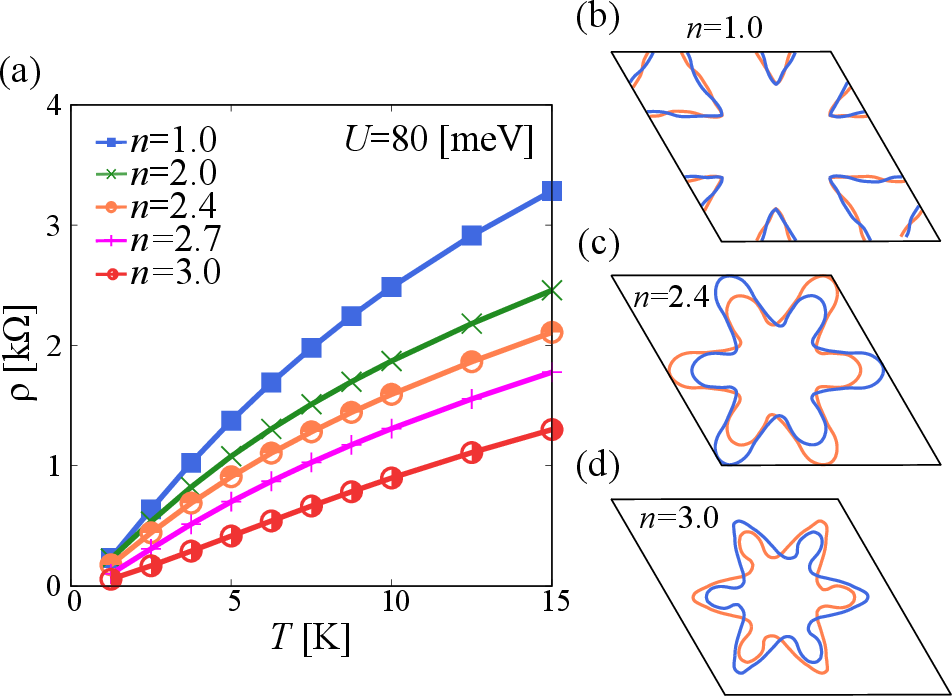}
	\caption{
	(a) $T$-dependence of $\rho$ for $n=1.0-3.0$.
	(b)-(d) FSs for $n=1.0, 2.4$ and $3.0$, respectively.
	}
	\label{fig:fig5}
	\end{figure}
Figure \ref{fig:fig5} shows the filling dependence of $\rho$, 
and the FSs for $n=1.0, 2.4$ and $3.0$.
The relation $\rho \propto T$ is satisfied in the various filling.
The fifteen-channel SU(4) fluctuations originate
from the (approximate) SU(4) symmetry which
the system possesses by nature in MATBG.
Thus, the SU(4) fluctuations easily
develop even away from vHS filling,
and therefore $\rho \propto T$ is realized in wide $n$ range.
Experimentally, $T$-linear resistivity is observed
in wide $n$ range \cite{Jaoui,Polshyn,Park}.
Thus, our results are consistent with experiments.
The $T$-linear resistivity realized in 
wide $n$ range suggests that
the SU(4) fluctuations universally develop and
non-Fermi liquid behavior in MATBG is mainly derived
from the SU(4) fluctuations.
The coefficient $a=\rho/T$ for $n=1.0$ is 
largest in $n=1.0-3.0$.
This filling dependence of the coefficient $a$
is similarly observed in experiments \cite{Jaoui,Polshyn,Park}.
The $n$-dependence of the
$\gamma_{\mathrm{cold}}$ is shown in Appendix C.
The obtained $\gamma_{\mathrm{cold}}$ is largest for $n=1.0$ 
due to the good nesting of the FS as shown in
Fig. \ref{fig:fig5}(b).

	\begin{figure}[!htb]
	  \includegraphics[width=.99\linewidth]{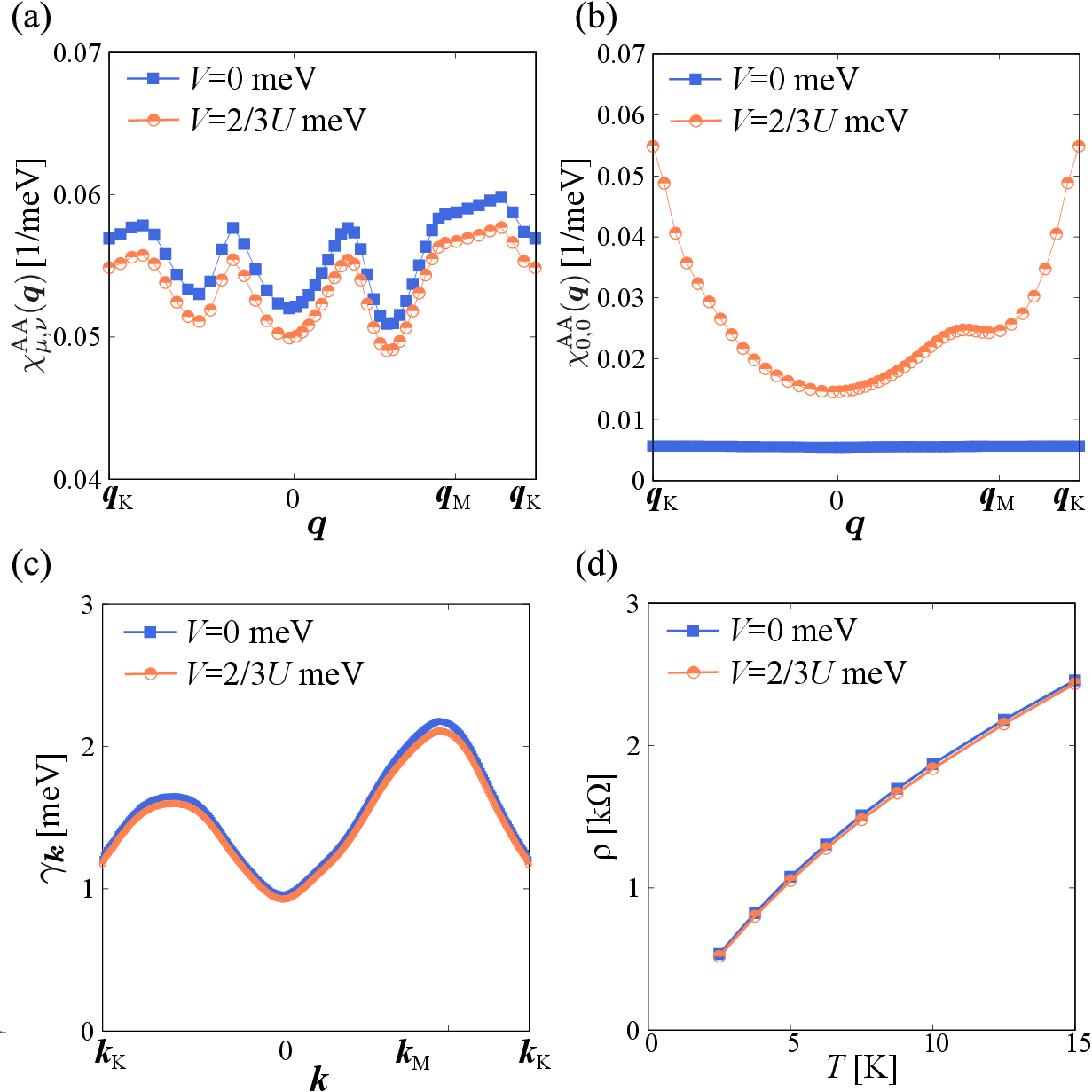}
	  \caption{
	  $\bm{q}$-dependences of 
	  (a) $\chi_{\mu, \nu}^{\mathrm{AA}}(\bm{q})$ 
	  except for $(\mu,\nu)=(0,0)$,
	  and 
	  (b) $\chi_{0,0}^{\mathrm{AA}}(\bm{q})$ for $V=0$ (blue line)
	   and $V_1=2/3U$ (orange line).
	   $\bm{q}_{\mathrm{K}}$ and $\bm{q}_{\mathrm{M}}$ are defined
	   in \ref{fig:fig3}(a).
	  (c) $\bm{k}$-dependences of $\gamma_{\bm{k}}$.
	  Here, $\bm{q}_{\mathrm{K}}$=$\left(\frac{2\pi}{\sqrt{3}}, -\frac{2\pi}{\sqrt{3}}\right)$
	  and $\bm{q}_{\mathrm{M}}$=$\left(\frac{2\pi}{\sqrt{3}}, 0\right)$
	  (d) $T$-dependence of $\rho$.}
	  \label{fig:S2}
	  \end{figure}


	  In metallic MATBG ($n$=1-3), 
	  the off-site Coulomb interaction is screened 
	  and becomes short ranged.
Here, we discuss the effect of the 
	  off-site Coulomb interaction based
	  on the Kang-Vafek model \cite{Vafek}.
	  We introduce the nearest-neighbor ($V_1$),
	  next nearest-neighbor ($V_2$), and
	  the third next nearest-neighbor ($V_3$)
	  hopping integral into the on-site
	  Coulomb interaction term in Eq. \ref{eqn:V}
	  Here, we fix $U=80\  \mathrm{meV}$, 
	  $V_1=2V_2=2V_3$ and $V_1=0$ or $V_1=2U/3$
	  The results given by the FLEX approximation 
	  for $V_1=0$ (blue line) and $V_1=2U/3$ 
	  are shown in Fig. \ref{fig:S2}.
	  Figure \ref{fig:S2}(a) shows the SU(4) susceptibility
	  $\chi_{\mu,\nu}^{\mathrm{AA}}(\bm{q})\ [(\mu,\nu)\neq(0,0)]$.
	  Although $\hat{\chi}_{\mu,\nu}(\bm{q})$ are slightly suppressed 
	  by the off-site Coulomb interaction, 
	  $\hat{\chi}_{\mu,\nu}(\bm{q})$ for $V_1=2U/3$ 
	  are fifteen-fold degenerated
	  and quantitatively unchanged.
	  In contrast, $\chi_{0,0}^{AA}(\bm{q})$ shown in Fig.
	  \ref{fig:S2}(b) is drastically changed
	  whether $V_1$ is zero or nonzero, and obtained 
	  $\hat{\chi}_{0,0}(\bm{q})$ for $V_1=2U/3$ 
	  is same order as $\chi_{\mu,\nu}(\bm{q})$.
	  By introducing the off-site Coulomb interactions,
	  the local charge susceptibility is modified
	  as 
	  \begin{align}
		\hat{\chi}_{0,0}(q) =\hat{\chi}_{0,0}^0(q)\left[\hat{1}+\left(\frac{3U}{4}+2\hat{v}(\bm{q})\right)
		\hat{\chi}^0_{0,0}(q)\right]^{-1}. 
	  \end{align}
	  Here, the formulation of $\hat{\chi}_{\mu,\nu}(q) 
	  \ [(\mu,\nu)\neq(0,0)]$ in Eq. (\ref{eqn:chi2}) is unchanged, because
	  the susceptibility $\hat{\chi}_{\mu,\nu;\mu',\nu'}
	  \ [(\mu',\nu')\neq(\mu,\nu)]$ is negligible.
	  Therefore, $\hat{\chi}_{0,0}(\bm{q})$ is only enlarged by $v_{\a\a'}$,
	  and other channels of the susceptibilities take almost the same value.
	  The obtained damping rate $\gamma_{\bm{k}}$ is shown in Fig. \ref{fig:S2}(c).
	  Nevertheless $\hat{\chi}_{0,0}(\bm{q})\thicksim \hat{\chi}_{\mu,\nu}(\bm{q})$
	  for $V_1 =2U/3$,
	  $\gamma_{\bm{k}}$ is almost equivalent to that
	  for $V_1=0$.
	  This is because the contribution of $\hat{\chi}_{0,0}(\bm{q})$
	  to $\gamma_{\bm{k}}$ is just 1/16 of other all channels,
	  and $\hat{\chi}_{\mu,\nu}(\bm{q})\ [(\mu,\nu)\neq(0,0)]$ is essentially independent of $V$.
	  Consequently, the resistivity $\rho$ obtained for
	  $V_1=2U/3$ is almost the same as that
	  for $V_1=0$.
	  Therefore, the present analysis based on the on-site
	  Coulomb interaction $U$ is justified.

\section{summary}

In this study, we demonstrated that the $T$-linear resistivity
is realized by the electron-electron correlation in MATBG
in the presence of the SU(4) valley+spin composite fluctuations.
We calculated the self-energy by employing the
FLEX approximation.
The obtained self-energy takes large value
due to the fifteen-fold degenerated SU(4) fluctuations.
Robust $T$-linear resistivity 
is realized for wide ranged $n$ at low temperatures
derived from the SU(4) fluctuations.
Importantly, the $T$-linear resistivity is realized
even when the
system is far from the SU(4) QCP 
($\a \lesssim 0.8$ in our calculations).
Then, large $T$-linear coefficient $a\equiv \rho/T$ is obtained 
in the present mechanism.
The $T$-linear coefficient $a$ due to only
the spin fluctuations is small, which is less than
1/10 of the coefficient observed in Ref. \cite{Jaoui,Polshyn}.
Thanks to the SU(4) fluctuations,
robust and large
$T$-linear resistivity is observed for wide $n$ range, 
even away from $n_{\mathrm{vHS}}=2.0$, consistent with experiments.
This result is strong evidence that the SU(4) fluctuations
universally develop in MATBG.

As well as MATBG, the exotic electronic states appear
in other twisted multilayer graphene.
For example, non-Fermi liquid type transport phenomena
\cite{Burg,Liu}
, unconventional superconductivity \cite{Shen,Liu,He}
, and nematic ordere \cite{Samajdar}
has been observed in twisted double bilayer graphene (TDBG).
Furthermore, in trilayer graphene, 
unconventional superconducting state appears \cite{Zhou}.
The present Green function formalism 
in the SU(4) symmetry limit
will be useful in analyzing the abovementioned problems.

\section{acknowledgements}
This study has been supported by Grants-in-Aid for Scientific
Research from MEXT of Japan (JP18H01175, JP20K03858, JP20K22328, JP22K14003),
and by the Quantum Liquid Crystal
No. JP19H05825 KAKENHI on Innovative Areas from JSPS of Japan.

\section{appendix A: FLEX approximation for multiorbital Hubbard models}

In this Appendix, we explain another foumulation of
the multiorbital FLEX approximation
based on the matrix expressions of the Coulomb interaction.
This method has been widely used for 
ruthenate \cite{Takimoto-FLEX}, 
cobaltates \cite{Yada},
Fe-based superconductors \cite{Kuroki,Kontani-FeSC,Onari-SCVC},
and heavy fermions \cite{Tazai-multipole1,Tazai-multipole2,Tazai-multipole3}.
It is confirmed that the formulation using SU(4) operator
developed in the main text
is equivalent with the following formulation.

The Coulomb interaction $H_U$ in Eq. (\ref{eqn:HU1}) is decomposed into spin and charge channel as
\cite{Kontani-rev2}
\begin{eqnarray}
H_U&=&\frac{U}{8} \sum_{i,\a}\sum_{\{\rho\},\{\xi\}}\left[ 
-\hat{\Gamma}^s_{\xi_1\xi_2,\xi_3\xi_4}(\hat{\bm\s}\otimes\hat{\bm\s})_{\rho_1\rho_2,\rho_3\rho_4} \right.
\nonumber \\
& &\left. -\hat{\Gamma}^c_{\xi_1\xi_2,\xi_3\xi_4} ({\hat\s}^0\otimes{\hat\s}^0)_{\rho_1\rho_2,\rho_3\rho_4} \right]
\nonumber \\
& &\times c_{i,\a\rho_1\xi_1}^\dagger c_{i,\a\rho_2\xi_2}c_{i,\a\rho_4\xi_4}^\dagger c_{i,\a\rho_3\xi_3},
\label{eqn:A-HU}
\end{eqnarray}
where $\hat{\bm{\s}}$ and $\hat{\s}^0$ are Pauli matrix and 
identity matrix, respectively
and $\xi_i$ is valley index.
Here, $\Gamma^s_{\xi_1\xi_2,\xi_3\xi_4}=U$ for $\xi_1=\xi_2=\xi_3=\xi_4$
and $\xi_1=\xi_3=-\xi_2=-\xi_4$,
and $\Gamma^s=0$ for others.
Also, 
$\Gamma^c_{\xi_1\xi_2,\xi_3\xi_4}=-U$ for $\xi_1=\xi_2=\xi_3=\xi_4$,
$\Gamma^c=-2U$ for $\xi_1=\xi_2=-\xi_3=-\xi_4$,
$\Gamma^c=U$ for $\xi_1=\xi_3=-\xi_2=-\xi_4$,
and $\Gamma^c=0$ for others.
The self-energy in the FLEX calculation is given as
\begin{align}
\Sigma_{\a\a'\xi}(k)&=\frac{T}{N}\sum_{q}G_{\a\a'\xi}(k-q)V_{\a\xi\xi',\a'\xi'\xi}(q),
\label{eqn:A-self}\\
V_{\a\xi\xi',\a'\xi'\xi}(q)&=\frac{U^2}{2}(3{\hat\Gamma}^s{\hat\chi}^s_{\a\a'}(q){\hat\Gamma}^s \nonumber \\
&\ \ \ \ \ \ \ \ \ \ +{\hat\Gamma}^c{\hat\chi}^c_{\a\a'}(q){\hat\Gamma}^c)_{\xi\xi',\xi'\xi},
\label{eqn:A-V} \\
{\chi}^0_{\a\xi_1\xi_2,\a'\xi_3\xi_4}(q)&\nonumber \\
=-\frac{T}{N}\sum_k &G_{\a\a'\xi_1}(k+q)G_{\a'\a\xi_2}(k)\delta_{\xi_1,\xi_3}\delta_{\xi_2,\xi_4},
\label{eqn:A-chi0}\\
{\hat\chi}^{s(c)}(q)&={\hat \chi}^0(q)({\hat 1}-{\hat\Gamma}^{s(c)}{\hat \chi}^0(q))^{-1},
\label{eqn:A-chi} 
\end{align}
where $\hat{\chi}^{s(c)}$ is the spin (charge) susceptibility 
\cite{Silva}.
The self-energy in the FLEX approximation is given by
solving Eqs. (\ref{eqn:A-self})-(\ref{eqn:A-chi}) self-consistently.
The coefficients for the self-energy
originated from the spin fluctuations and
the charge fluctuation are 3/2 and 1/2, respectively.
The spin (charge) Stoner factor $\a^{s(c)}$ is defined as
the maximum eigenvalue of ${\hat \Gamma}^{s(c)}\hat{\chi}^{s(c)}_{\a\a'}$.
$\a^s$ and $\a^c$ are exactly equivalent due to the relation $U=U'$.

In the presence of the off-site Coulomb interaction
between $(i,\a)$ and $(j,\a')$, $v_{i\a,j\a'}$
given as Eq. (\ref{eqn:off-site}),
the effect of off-site Coulomb interaction in the FLEX approximation 
is simply given by replacing ${\hat \Gamma}^c$ with  
${\hat \Gamma}^c+v_{\a\a'}(\q)\delta_{\xi_1,\xi_2}\delta_{\xi_3,\xi_4}$
in Eqs. (\ref{eqn:A-V}) and (\ref{eqn:A-chi}).
Here, $v_{\a\a'}(\q)$ is the Fourier transform of $v_{i\a,j\a'}$.

The SU(4) susceptibility in Eq. (\ref{eqn:chi2}) can be expanded by
the spin and charge susceptibilities in Eq. (\ref{eqn:A-chi}) as
\begin{equation}
	\begin{split}
		\chi_{\mu,\nu}^{\a\a'}(\q,i\w_l)&=\int_0^\beta d\tau \langle O_{\mu,\nu}^{\a}(\q,\tau)O_{\mu,\nu}^{\a'}(-\q,0)\rangle
e^{i\w_l\tau}\\
&=\sum_{l_1l_2l_3l_4} Q^{\mu,\nu}_{\a l_1 l_2} \chi_{\a l_1 l_2,\a' l_3 l_4}(q)Q^{\mu,\nu}_{\a' l_3 l_4},
\label{eqn:A-chi2}
	\end{split}
\end{equation}
where ${Q}^{\mu,\nu}_{\a l l'}=({\hat \s}_\mu \otimes{\hat \tau}_\nu)_{l l'}$
and $l_i=(\rho_i,\xi_i)$.
The general susceptibility in the right-hand-side of 
Eq. (\ref{eqn:A-chi2}) is given as
\begin{align}
\chi_{\a l_1 l_2,\a' l_3 l_4}(q)&=\frac12\chi_{\a\xi_1\xi_2,\a'\xi_3\xi_3}^s(q)(\hat{\bm\s}\otimes\hat{\bm\s})_{\rho_1 \rho_2,\rho_3 \rho_4}
\nonumber \\
&+\frac12\chi_{\a \xi_1 \xi_2,\a' \xi_3 \xi_3}^c(q)({\hat\s}^0\otimes{\hat\s}^0)_{\rho_1 \rho_2,\rho_3 \rho_4}.
\label{eqn:A-chi3}
\end{align}
This conventional formalism used in Refs. 
\cite{Takimoto-FLEX,Yada,Kuroki,Kontani-FeSC,Onari-SCVC,Tazai-multipole1,Tazai-multipole2,Tazai-multipole3}
is exactly equivalent
with the SU(4) operator formalism explained in the main text.

\section{appendix B: resistivity within the second-order perturbation theory}
\begin{figure}[!htb]
	\includegraphics[width=.99\linewidth]{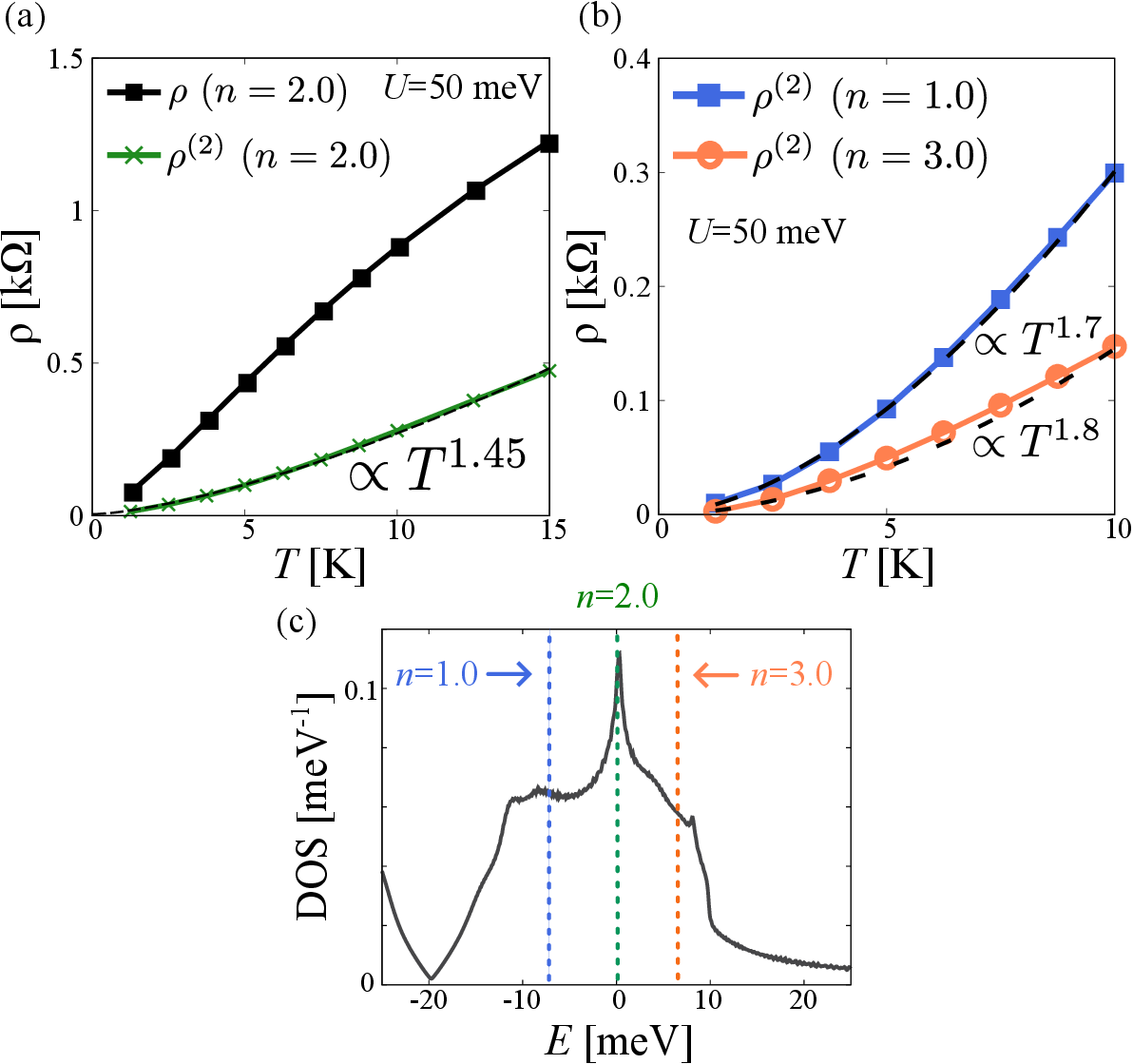}
	\caption{
	(a) $T$-dependence of $\rho$ obtained by the FLEX approximation 
	with full-order (black line) 
	and that in the self-consistent second-order perturbation theory (green line)
	for $n=2.0$.
	(b) $T$-dependence of $\rho$ obtained by the
	self-consistent second-order perturbation theory
	for $n=1.0$ (blue line) and $n=3.0$ (orange line).
	(c) The total DOS, broken lines with blue, green and orange
	represent the Fermi energy for $n=1.0$, $n=2.0$ and $n=3.0$,
	respectively.
	}
	\label{fig:A1}
	\end{figure}
Here, we discuss the important effect of vHS points on
the resistivity $\rho$ in the weak coupling region.
In the main text, 
the obtained power $m$ in $\rho=aT^m$
is smaller than about 1.5 for $n=2.0$, 
even in the case of very weak on-site Coulomb interaction $U$.
This result is inconsistent with
the expected behavior that 
the Fermi liquid behavior $\rho \propto T^2$ 
is obtained for the limit $U\rightarrow0$.
To understand this inconsistence,
we calculate the resistivity $\rho^{(2)}$,
which is given by the self-consistent second-order perturbation theory
with respect to $U$.
Figure \ref{fig:A1} (a) shows $\rho$ with full order (black line)
and within second-order perturbation theory (green line) 
with respect to $U$,
and (b) shows $\rho^{(2)}$ for $n=1.0$ (blue line)
and $n=3.0$ (orange line).
We set $U=50$ meV in Fig. \ref{fig:A1}.
The obtained power $m$ in $\rho^{(2)}=aT^m$ for $n=2.0$
is $m=1.45$, and
this is almost same with $m$ for $U=12.5$ meV in
Fig. \ref{fig:fig4}.
In contrast, the power $m$ in $\rho^{(2)}$ for
$n=1.0, 3.0$ are close to 2.
This results suggest that the power $m$
is enhanced by the effect of vHS points 
and $T$-linear resistivity is easily realized
near the vHS points.

\section{Appendix C: Filling dependence of $\gamma_{\mathrm{cold}}$}
\begin{figure}[!htb]
	\includegraphics[width=.55\linewidth]{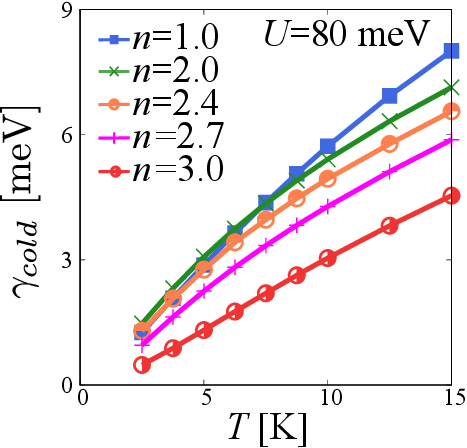}
	\caption{(a) $T$-dependence of the damping rate
	at cold spot, $\gamma_{\mathrm{cold}}$
	for $n=1.0-3.0$.
	}
	\label{fig:A2}
	\end{figure}
Figure \ref{fig:A2} shows 
the filling dependence of $\gamma_{\mathrm{cold}}$
for $n=1.0-3.0$.
$\gamma_{\mathrm{cold}}$ for $n=2.0-3.0$ get small
as the filling is far from $n_{\mathrm{VHS}}\simeq 2.0$.
Unexpectedely, the obtained $\gamma_{\mathrm{cold}}$ for $n=1.0$ at $T\gtrsim 10$K
takes larger value than it for $n=2.0$
in our calculation.
The reason is that the nesting condition on the FS for $n=1.0$
in Fig. \ref{fig:fig5} (b)
is better than that for $n=2.0$ in Fig. \ref{fig:fig1} (b).
Consequently, SU(4) susceptibilities for $n=1.0$ 
are higher than that for $n=2.0$ by reflecting the 
good nesting condition of the FS. (FS for $n=1.0$ is
shown in Fig. \ref{fig:fig5} (b).)
Thus, $\gamma_{\mathrm{cold}}$ for $n=1.0$ takes the largest value
due to the stronger nesting effect,
which exceeds the effect of the reduced DOS.


\end{document}